\pdfoutput=1

\documentclass[11pt,twoside,a4paper,cmspaper,final,collab]{cms-tdr}
\def\svnVersion{427520P}\def\cmsCernNoTag{CERN-EP-2016-308}\def\cmsCernDate{\today}\def\cmsMessage{Published in Physics Letters B as \href{http://dx.doi.org/10.1016/j.physletb.2017.04.071}{\doi{10.1016/j.physletb.2017.04.071.}}}

\begin{document}\cmsNoteHeader{SMP-14-018}

\hyphenation{had-ron-i-za-tion}
\hyphenation{cal-or-i-me-ter}
\hyphenation{de-vices}
\RCS$Revision: 403055 $
\RCS$HeadURL: svn+ssh://svn.cern.ch/reps/tdr2/papers/SMP-14-018/trunk/SMP-14-018.tex $
\RCS$Id: SMP-14-018.tex 403055 2017-05-08 03:38:01Z zhaoru $
\newlength\cmsFigWidth
\ifthenelse{\boolean{cms@external}}{\setlength\cmsFigWidth{0.98\columnwidth}}{\setlength\cmsFigWidth{0.85\textwidth}}
\ifthenelse{\boolean{cms@external}}{\providecommand{\cmsLeft}{top\xspace}}{\providecommand{\cmsLeft}{left}}
\ifthenelse{\boolean{cms@external}}{\providecommand{\cmsRight}{bottom\xspace}}{\providecommand{\cmsRight}{right}}

\newcommand{\jj}{\ensuremath{\mathrm{jj}}\xspace}

\cmsNoteHeader{SMP-14-018}

\title{Measurement of the cross section for electroweak production of \texorpdfstring{$\Z\gamma$}{Z gamma} in association with two jets and
constraints on anomalous quartic gauge couplings in proton-proton collisions at \texorpdfstring{$\sqrt{s} = 8\TeV$}{sqrt(s) = 8 TeV}}

\date{\today}

\abstract{
A measurement is presented of the cross section for the electroweak production of
a Z boson and a photon in association with two jets
in proton-proton collisions at $\sqrt{s} = 8\TeV$.
The Z~bosons are identified through their decays to electron or muon pairs.
The measurement is based on data collected with the CMS detector corresponding to an integrated luminosity of 19.7\fbinv.
The electroweak contribution has a significance of 3.0 standard deviations,
and the measured  fiducial cross section is  $1.86_{-0.75}^{+0.90}\stat_{-0.26}^{+0.34}\syst\pm 0.05\lum$\unit{fb}, while the summed electroweak
and quantum chromodynamic total cross section in the same region is observed to be
$5.94_{-1.35}^{+1.53}\stat_{-0.37}^{+0.43}\syst\pm 0.13\lum$\unit{fb}.
Both measurements are consistent with the leading-order standard model predictions.
Limits on anomalous quartic gauge couplings are set
based on the $\Z\gamma$ mass distribution.}

\hypersetup{%
pdfauthor={CMS Collaboration},%
pdftitle={Measurement of the cross section for electroweak production of Z gamma in association with two jets and
constraints on anomalous quartic gauge couplings in proton-proton collisions at sqrt(s) = 8 TeV},%
pdfsubject={CMS},%
pdfkeywords={CMS, physics, aQGC, electroweak production}}

\maketitle
\section{Introduction}
\label{sec:intro}

With the discovery of the Higgs boson at the CERN LHC~\cite{Aad:2012tfa,Chatrchyan:2012ufa}, the standard model (SM) became a great success.
The high energy and luminosity of the LHC provides the opportunity to observe many processes that are predicted by the SM,
including electroweak production of multiple gauge bosons
($\PW\mathrm{V}\gamma$~\cite{Chatrchyan:2014bza}, V$\gamma\gamma$~\cite{Aad:2015uqa,Aad:2016sau,CMS-PAS-SMP-15-008}),
vector boson scattering (VBS) (same charge $\PW^\pm\PW^\pm$ scattering~\cite{Aad:2014zda,Aaboud:2016ffv,CMS-PAS-SMP-13-015},
$\gamma\gamma\to\PWp\PWm$~\cite{Khachatryan:2016mud}, EW $\PW\gamma\jj$~\cite{Khachatryan:2016vif}, $\PW^\pm\Z$~\cite{Aad:2016ett}),
and vector boson fusion (VBF) (EW W(Z)jj~\cite{Chatrchyan:2013jya,CMS-PAS-FSQ-12-035,Aad:2014dta,CMS-PAS-SMP-13-012}).
Same charge $\PW^\pm\PW^\pm$ scattering has been observed by ATLAS, and the exclusive $\gamma\gamma\to\PWp\PWm$ process by CMS, both with significances larger than 3 standard deviations.
The triboson final state $\Z\gamma\gamma$ has been observed by ATLAS and CMS with a significance larger than 5 standard deviations.
The EW production of a \Z boson (decaying into two oppositely-charged leptons), a photon, and two jets (henceforth denoted $\cPZ\gamma\jj$) has never been studied before, and is the subject of this paper.
While the cross section for quantum chromodynamic (QCD) induced $\cPZ\gamma\jj$ production is orders of magnitude larger than the one for EW production, the latter can be used
to perform important tests of the SM, and to search for contributions from physics beyond the SM that could manifest themselves as anomalous trilinear or quartic gauge boson couplings (aTGC or aQGC~\cite{Chatrchyan:2014bza,Aad:2015uqa,Aad:2016sau,CMS-PAS-SMP-15-008,Aad:2014zda,CMS-PAS-SMP-13-015,Khachatryan:2016mud,CMS-PAS-SMP-14-011,Aad:2016ett}).

\begin{figure*}[h!tb]
\centering
\includegraphics[width=1\linewidth]{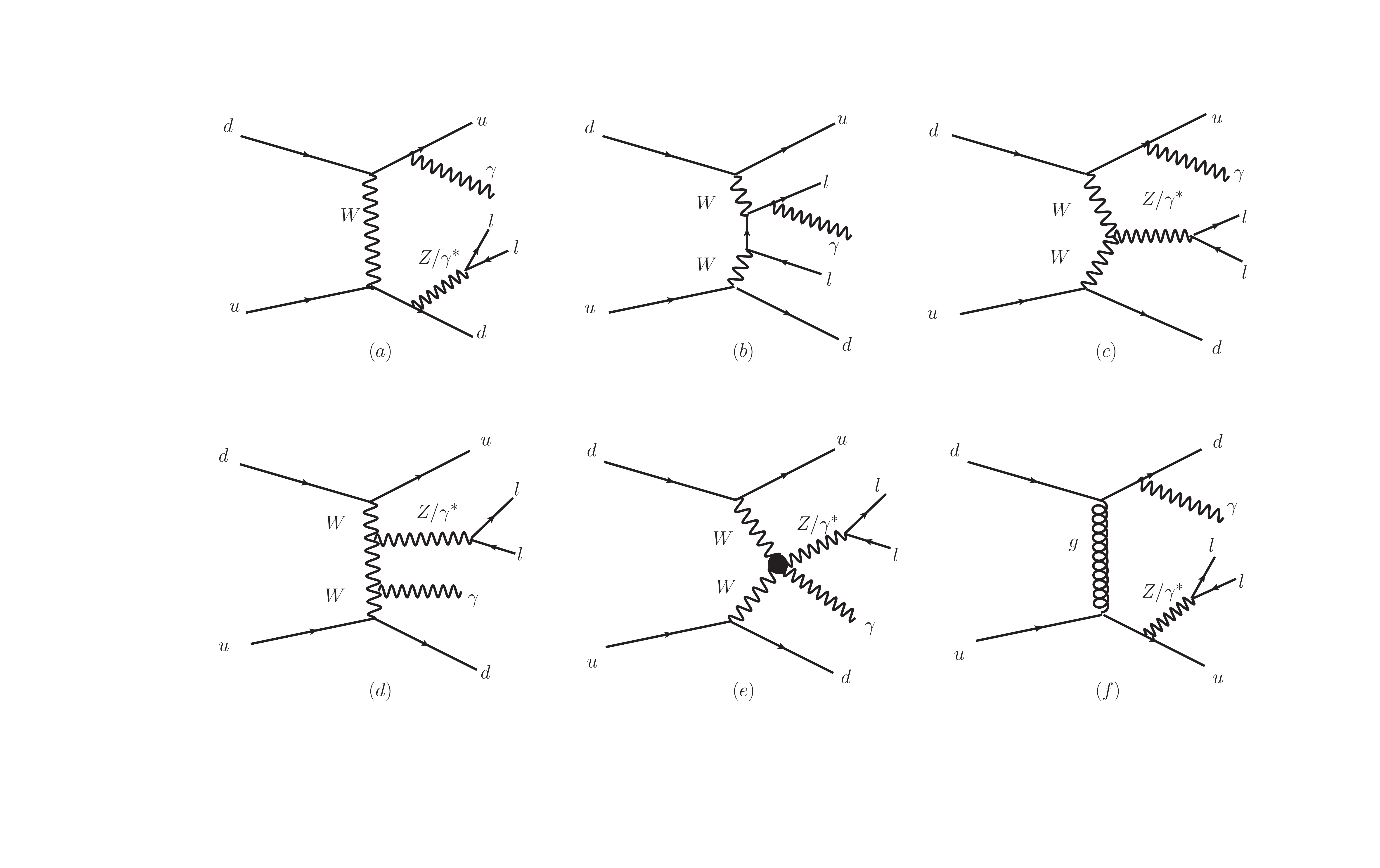}
\caption{Representative diagrams for EW $\cPZ\gamma\jj$ production at the LHC: (a) bremsstrahlung, (b) multiperipheral, (c,d) VBF with TGC, (e) VBS including QGC, and (f) Example diagram for the QCD $\cPZ\gamma\jj$ production.}
\label{fig:za_feynman}
\end{figure*}

This letter presents a measurement of the associated EW production of $\cPZ\gamma$jj,
using the 8\TeV proton-proton collision data recorded by the CMS detector.
The major processes contributing to EW $\cPZ\gamma\jj$ production are represented by the Feynman diagrams in Fig.~\ref{fig:za_feynman}. They are (a)~bremsstrahlung, (b)~multiperipheral (or non-resonant) production, (c,d)~VBF with either two trilinear gauge boson couplings (TGC), or (e)~VBS with quartic gauge boson couplings (QGC). The VBS processes are particularly interesting because they involve QGCs (e.g. $\PW\PW\Z\gamma$).
It is not possible, however, to isolate the QGC processes from the other contributions, such as the double TGC processes that are topologically similar.
The interference of the VBS diagrams ensures unitarity of the VBS cross section in the SM at high energy.
We present measurements of the combined cross sections for all EW processes that result in the $\cPZ\gamma\jj$ final state.
The main background source is $\cPZ\gamma\jj$ production where the associated jets are produced through QCD-induced processes (such as the Feynman diagram given in Fig.~\ref{fig:za_feynman}(f)).
Other backgrounds include jets or leptons misidentified as photons,
diboson processes in which a W or Z boson decays into two jets and the photon originates from initial or final-state radiation, and contributions from top quark pairs and single top quark production.

\section{The CMS detector }
\label{sec:CMSdet}

The central feature of the CMS apparatus is a superconducting solenoid of 6\unit{m} internal diameter,
providing a magnetic field of 3.8\unit{T}.
Within the solenoid volume are a silicon pixel and strip tracker,
a lead tungstate crystal electromagnetic calorimeter (ECAL),
and a brass and scintillator hadron calorimeter (HCAL),
each composed of a barrel and two endcap sections.
Forward calorimeters extend the pseudorapidity, $\eta$,
coverage provided by the barrel and endcap detectors.
Muons are measured in gas-ionization detectors embedded in the steel flux-return yoke outside the solenoid.

The particle-flow (PF) event algorithm~\cite{CMS-PAS-PFT-09-001,CMS-PAS-PFT-10-001} reconstructs and identifies each individual particle with an optimized combination of information from the various elements of the CMS detector. The energy of photons is directly obtained from the ECAL measurement, corrected for zero-suppression effects. The energy of electrons is determined from a combination of the electron momentum at the primary interaction vertex as determined by the tracker, the energy of the corresponding ECAL cluster, and the energy sum of all bremsstrahlung photons spatially compatible with originating from the electron track. The energy of muons is obtained from the curvature of the corresponding track. The energy of charged hadrons is determined from a combination of their momentum measured in the tracker and the matching ECAL and HCAL energy deposits, corrected for zero-suppression effects and for the response function of the calorimeters to hadronic showers. Finally, the energy of neutral hadrons is obtained from the corresponding corrected ECAL and HCAL energy.

In the barrel section of the ECAL, an energy resolution of about 1\% is achieved for unconverted or late-converting photons in the tens of \GeVns energy range.
The resolution for other photons in the barrel section is about 1.3\% up to $\abs{\eta}=1$, rising to about 2.5\% at $\abs{\eta}=1.4$.
In the endcaps, the resolution for unconverted or late-converting photons is about 2.5\%, and the resolution for the remaining photons in the endcap is between 3\% and 4\%~\cite{CMS:EGM-14-001}.
When combining information from the entire detector, the jet energy resolution is typically 15\% at 10\GeV, 8\% at 100\GeV, and 4\% at 1\TeV.

Muons are measured in the range of $\abs{\eta}< 2.4$, with detection planes utilizing three technologies: drift tubes, cathode strip chambers, and resistive-plate chambers.
Matching muons to tracks measured in the silicon tracker results in a \pt resolution for muons with $20 <\pt < 100\GeV$ of 1.3--2.0\% in the barrel and better than 6\% in the endcaps.

The electron momentum is estimated by combining the energy measurement in the ECAL with the momentum measurement in the tracker. The momentum resolution for electrons with transverse momentum
$\pt\approx45\GeV$ from $\Z \to \Pe \Pe$ decays ranges from 1.7\% for nonshowering electrons in the barrel region to 4.5\% for showering electrons in the endcaps.
The dielectron mass resolution for $\Z \to \Pe \Pe$ decays is 1.9\% when both electrons are in the ECAL barrel, and 2.9\% when both electrons are in the endcaps.

A more detailed description of the CMS detector, together with a definition of the coordinate system used and the relevant kinematic variables, can be found in Ref.~\cite{Chatrchyan:2008zzk}.

\section{Event reconstruction and selection}
\label{sec:eventselec}

Candidate events are selected online with triggers that require two muons or electrons, where the leading and subleading leptons have $\pt>17$ and 8\GeV respectively, with $\abs{\eta} < 2.4$ (muons) or $\abs{\eta} < 2.5$ (electrons).
The overall trigger efficiency is about 94\% and 90\% for muons and electrons, respectively,
with a small dependence on \pt and $\eta$.

Muons are reconstructed with a global fit using both the inner tracking system and the muon spectrometer. An isolation requirement is applied in order to suppress the background from multijet events~\cite{Chatrchyan:2012xi,Chatrchyan:2015jya}.
Electron candidates are reconstructed by matching energy deposits in the ECAL with reconstructed tracks; they must pass stringent quality criteria and an isolation requirement~\cite{Chatrchyan:2013dga}.
Charged leptons  must originate from the primary vertex,
 which is defined as the vertex whose tracks have the highest sum of $\pt^2$.
We require that each event has exactly two oppositely charged muons (electrons) with $\pt> 20$\GeV and $\abs{\eta}<2.4\,(2.5)$ and that the invariant mass of the dilepton system must satisfy $70 < M_{\ell\ell} <110\GeV$. The selection efficiencies for leptons are measured using the tag-and-probe method~\cite{VBTF} and are approximately 96\% for the muons ~\cite{Khachatryan:2015hwa} and 80\% for the electrons~\cite{Chatrchyan:2012xi}.

Photon candidates are reconstructed from energy deposits in the ECAL with no associated track. Quality selection criteria~\cite{CMS:EGM-14-001} are applied to the reconstructed photons to suppress the background from hadrons misidentified as photons.
The observables used in the photon selection are: (1)~PF-based isolation variables that are corrected for the contribution from additional proton-proton collisions in the same bunch crossing (pileup); (2)~a small ratio of hadronic energy in the HCAL to electromagnetic energy in the ECAL matched in $(\eta,\phi)$ (where $\phi$ is azimuthal angle in radians); (3)~the transverse width of the electromagnetic shower along the $\eta$ direction ~\cite{CMS:EGM-14-001}; and (4)~an electron track veto.
We consider only photons in the ECAL barrel region ($\abs{\eta}<1.44$) with $\pt>25\GeV$. Events with the photon candidate in one of the endcaps ($\abs{\eta}>1.57$) are excluded from the selection because their signal purity is lower and systematic uncertainties are large.

Hadronic jets are formed from the particles reconstructed by the PF algorithm, using the \FASTJET software package~\cite{Cacciari:2011ma} and the anti-\kt jet clustering algorithm~\cite{Cacciari:2008gp} with a distance parameter of 0.5.
To reduce the contamination from pileup,
charged PF candidates in the tracker acceptance region $\abs{\eta}<2.4$, are excluded from the jet clustering procedure if associated with pileup vertices.
The contribution of neutral particles from pileup events to the jet energy is taken into account by means of a correction based on the projected area of the jet on the front face of the calorimeter.
Jet energy corrections are derived from a measurement of the \pt balance in
dijet and photon+jet events in data~\cite{Chatrchyan:2011ds}. Further residual corrections as functions of \pt and $\eta$ are applied to the data to correct for the small differences between data and simulation. Additional quality criteria are applied to the jets in order to remove spurious jet-like features originating from isolated noise patterns in the calorimeters or in the tracker~\cite{CMS-PAS-JME-10-003}.
The two jets with the highest \pt are tagged as the signal jets and are required to have $\pt> 30$\GeV and $\abs{\eta}< 4.7$.
Since we are primarily interested in the VBS topologies, we require that the invariant mass of the two jets, $M_\jj>150\GeV$.

Table~\ref{tab:selection} presents a summary of the three different section criteria that are used for (1)~the SM EW signal search, (2)~the SM fiducial cross section measurement, and (3)~the aQGC searches.
The criteria isolate events consistent with the VBS topology of two high-energy scattered jets separated by a large rapidity gap. The cross section measurement adds two variables sensitive to the VBS process: $\abs{y_{\Z\gamma}-(y_{\mathrm{j}1}+y_{\mathrm{j}2})/2}$, which ensures the $\Z\gamma$ systems is located between the scattered jets in eta; and $\Delta\phi_{\Z\gamma,\jj}$, which requires the $\Z\gamma$ system transverse momentum is consistent with recoiling against the transverse momentum of the two combined jets.
The fiducial cross section criteria constrain the VBS topology with only basic kinematic cuts that define the acceptance of the CMS detector and a simple two dimensional requirement on the rapidity separation and invariant mass of the jets.
A tight $\pt^{\gamma}$ selection is applied to reach a higher expected significance in a search for a possible aQGC signal
in the EW $\Z\gamma\jj$ process.

\begin{table*}[htb]
\centering
\topcaption{Summary of the three different event criteria: (1) selection for the EW signal measurement; (2) the cross section measurement; and (3) the selection for the aQGC search. ``j1" and ``j2" represent the jets that have the largest and second-largest $\pt$, ``$\ell1$" and ``$\ell2$" denote the lepton and antilepton from the decay of the Z boson, $y$ is the rapidity, $\Delta\phi_{\Z\gamma,\jj}$ is the absolute difference between $\phi_{\Z\gamma}$ and $\phi_{\mathrm{j1j2}}$, and the angular separation $\Delta R = \sqrt{\smash[b]{(\Delta\eta)^{2}+(\Delta\phi)^{2}}}$.}
\begin{tabular}{ccc}
\hline
\multicolumn{3}{c}{Common selection}   \\
 \hline
\multicolumn{3}{c}{$\pt^{\mathrm{j1,j2}}> 30$\GeV, $\abs{\eta^{\mathrm{j1,j2}}}<4.7$} \\
\multicolumn{3}{c}{$\pt^{\mathrm{\ell1,\ell2}} > 20\GeV$, $\abs{\eta^{\ell1,\ell2}}< 2.4$} \\
\multicolumn{3}{c}{$\abs{\eta^{\gamma}} < 1.4442$} \\
\multicolumn{3}{c}{$M_{\jj}>150\GeV$}  \\
\multicolumn{3}{c}{$70 <M_{\ell\ell}<110\GeV$}  \\[2ex]
\hline
EW signal measurement                                   & Fiducial cross section              & aQGC search           \\
  \hline
  $\pt^{\gamma} > 25\GeV $                                           & $\pt^{\gamma} > 20$\GeV              & $\pt^{\gamma} > 60$\GeV \\
  $\abs{\Delta\eta_{\jj}}>1.6$                                       & $|\Delta\eta_{\jj}|>2.5$         & $|\Delta\eta_{\jj}|>2.5$\\
$\Delta R_{\mathrm{j}\ell}>0.3$, $\Delta R_{{\jj}, \gamma{\mathrm{j}}, \gamma{\ell}}>0.5$ & $\Delta R_{{\jj}, \gamma{\rm{j}}, \gamma{\ell}, {\mathrm{j}\ell}}>0.4$ & $\Delta R_{\mathrm{j}\ell}>0.3$, $\Delta R_{{\jj}, \gamma{\mathrm{j}}, \gamma{\ell}}>0.5$  \\
$\abs{y_{\Z\gamma}-(y_{\mathrm{j}1}+y_{\mathrm{j}2})/2}<1.2$  & $M_{\jj}>400$\GeV          &  $M_{\jj}>400$\GeV       \\
  $\Delta\phi_{\Z\gamma,\jj}>2.0$ radians                                 &                                       &      \\
  $M_{\jj}>400$\GeV with two divided regions           &    &    \\
 400 $<M_{\jj}<800$\GeV and $M_{\jj}>800$\GeV  &  &  \\
 \hline
  \end{tabular}
  \label{tab:selection}
\end{table*}
\section{Data and simulation}
\label{sec:samples}

We use data collected with the CMS detector, corresponding to an integrated luminosity of 19.7\fbinv, at proton-proton center-of-mass energy of 8\TeV.

The EW signal, $\Z\gamma\jj$, at leading-order (LO), and the main background, QCD $\Z\gamma$ with 0--3 additional jets, for which the next-to-leading-order (NLO) QCD prediction has been taken from Ref.~\cite{Campanario:2014wga}, matched with parton shower based on the so-called ``MLM prescription"~\cite{mlm, Alwall:2007fs}, are simulated using \MADGRAPH v5.1.3.30~\cite{Alwall:2011uj} interfaced with \PYTHIA v6.424~\cite{Sjostrand:2006za} for hadronization and showering, using a CTEQ6L1 parton distribution function (PDF) set~\cite{Pumplin:2002vw}.
The second significant background  contribution comes from processes where a jet is misidentified as a photon (fake photon), and this contribution is estimated from data.
Other background contributions come from diboson processes (WW/WZ/ZZ) simulated by \PYTHIA , single top processes simulated by \POWHEG ,
and $\ttbar\gamma$ simulated using \MADGRAPH interfaced with \PYTHIA.
The next-to-leading-order QCD cross sections are used to normalize these simulated samples, except for $\ttbar\gamma$ where an LO prediction is taken.

All the simulated events are processed through a \GEANTfour~\cite{geant4} simulation of the CMS detector.
The tag-and-probe technique is used to correct for
data-Monte Carlo (MC) differences in the trigger efficiency, as well as the reconstruction and selection efficiencies.
Additional proton-proton interactions are superimposed over the hard
scattering interaction with the distribution of primary vertices matching that
obtained from the collision data.
\section{Background modeling}
\label{sec:qcdmod}
The dominant source of background to the EW signal is QCD $\Z\gamma\text{ + jets}$ production.  The shape of this background is taken from MC simulation and the normalization is evaluated from data in a control region, defined as $150 <M_{\jj}<400$\GeV, where the signal contribution is below 1\%.
The simulated MC events correctly reproduce the yield of these events with a correction factor of 1.00 $\pm$ 0.22 for the combined  $\Z\to\Pgmp\Pgmm$ and $\Z\to\Pep\Pem$ channels. The value is comparable with the NLO QCD $K$ factor from Ref.~\cite{Campanario:2014wga}, which is around 1.1 for $M_{\jj}<400$\GeV.

The background from fake photons arises mainly from
Z+jets events where one jet satisfies the photon ID criteria.
The estimation is based on events similar to the ones selected with the baseline selection described in Table~\ref{tab:selection}, except that the photon must fail the tight photon ID and satisfy a looser ID
requirement based on the charged isolation variable.
This selection ensures that the photon arises from a jet, but
still has kinematic properties similar to a genuine photon satisfying the tight photon ID.
We select genuine photons using $\sigma_{\eta \eta}$, a photon identification variable that exploits the small lateral extension of the electromagnetic shower~\cite{Khachatryan:2015hwa,CMS:EGM-14-001}.
Based on the difference between the $\sigma_{\eta \eta}$ distributions for fake photons and genuine photons, a fit is made
to normalize the number of events with fake photons to the number of events with genuine photons and obtain the probability to have a fake photon.
The fake photon probability is calculated based on different ${\pt^{\gamma}}$ regions in a manner similar to that described in Ref.~\cite{Vgamma}.

Other backgrounds, including top quark and diboson production processes are estimated from MC simulations and normalized to the integrated luminosity of the data sample.
The contribution from these backgrounds is less than 10\% after applying the kinematic selection (Section~\ref{sec:eventselec}) and is negligible once the final EW and aQGC selection criteria (Sections~\ref{sec:sigif} and~\ref{sec:aqgcparam}) are applied.

\begin{figure}[h!tb]
\centering
         \includegraphics[width=0.49\textwidth]{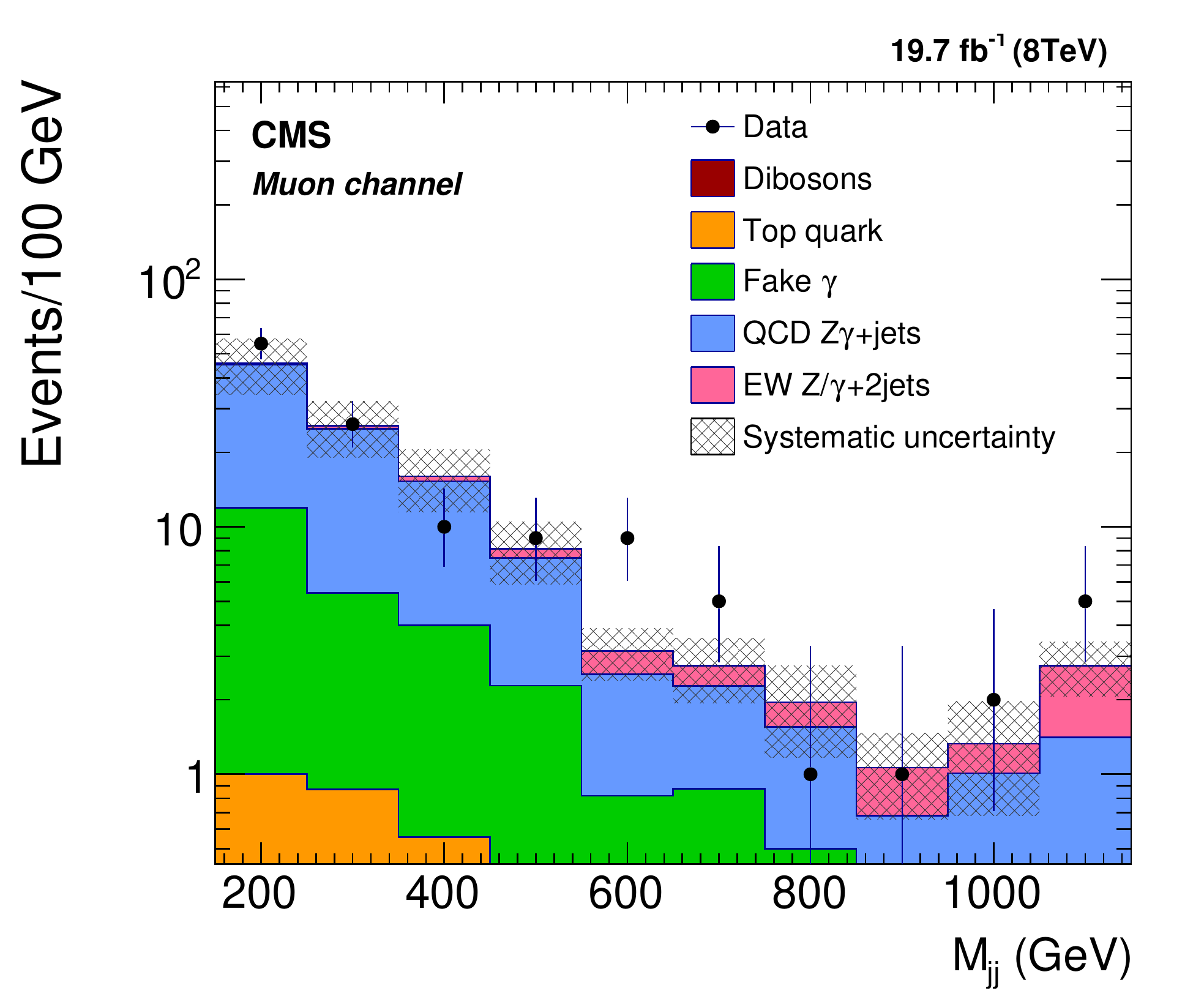}
         \includegraphics[width=0.49\textwidth]{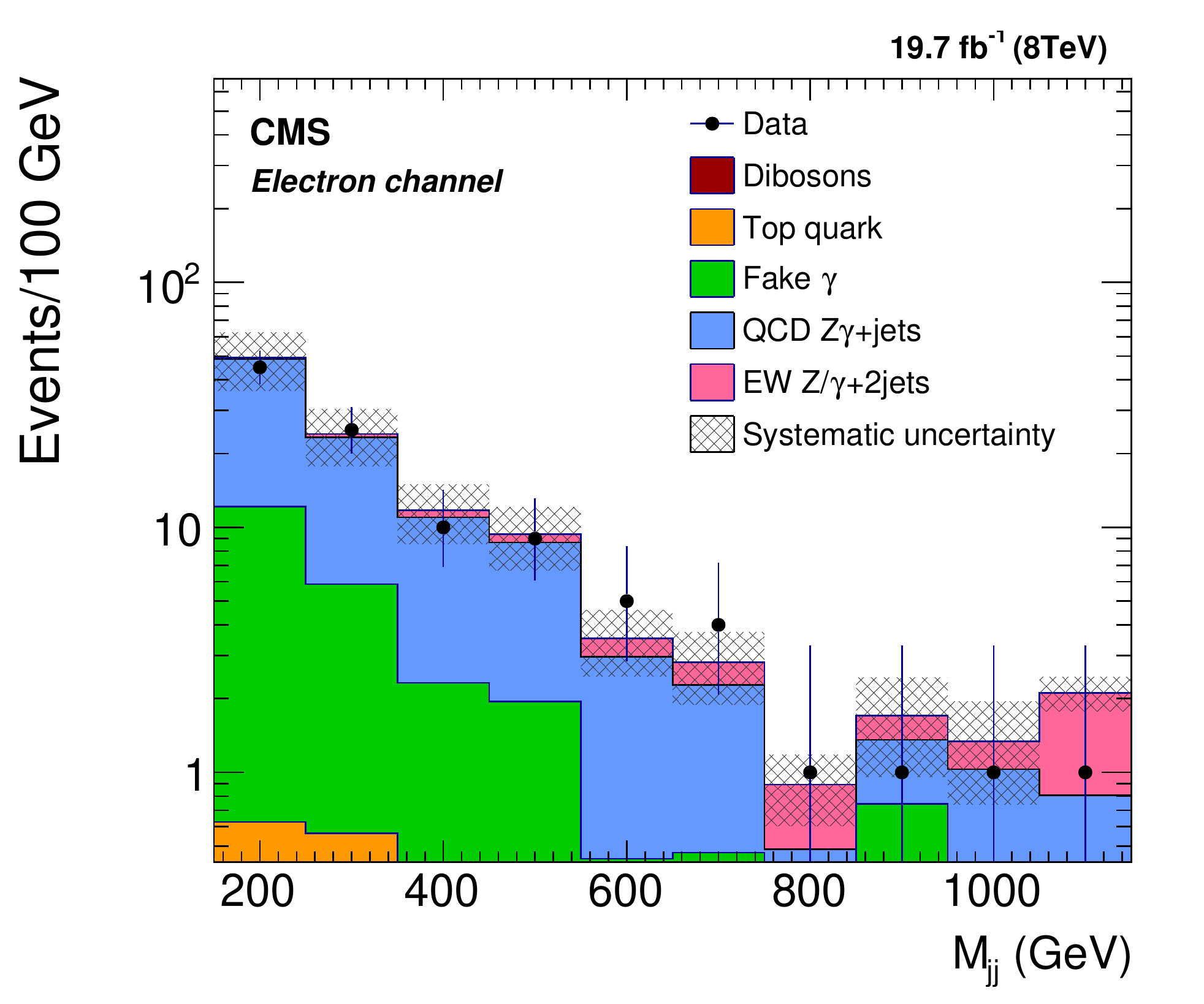}
         \caption{The $M_{\jj}$ distributions measured in (\cmsLeft) muon and (\cmsRight) electron channels.
The data (solid symbols with error bars representing the statistical uncertainties) are compared to a data-driven background estimate, combined with MC predictions for the signal contribution.
The hashed bands represent the full uncertainty in the predictions, as described in Section~\ref{sec:sysunc}. The last bin includes overflow events.
         \label{fig:MJJ}}
\end{figure}

The $M_{\jj}$ distributions for the $\Z\to\Pgmp\Pgmm$ and $\Pep\Pem$ channels after the selection
requirements described in Section~\ref{sec:eventselec} are shown in Fig.~\ref{fig:MJJ}.
The observed distributions are compared to the combined prediction of the backgrounds and of the EW $\Z\gamma\jj$ signal.

\section{Systematic uncertainties}
\label{sec:sysunc}

The systematic uncertainty in the QCD $\Z\gamma$+jets background estimation is 22\% for both $\Z\to\Pgmp\Pgmm$ and $\Z\to\Pep\Pem$; it is dominated by the large statistical uncertainty in the control region used for normalization.
The shape uncertainties that are related to the extrapolation of the normalization factor to the signal region ($M_{\jj}>400$\GeV) are determined by varying the renormalization and factorization scales as well as the MLM matching scale~\cite{mlm, Alwall:2007fs} up and down by a factor of two.
Finally, we combine both the normalization factor uncertainty and the shape uncertainty to obtain the total uncertainty.

The systematic uncertainty in the background estimation from fake photons arises from
the variation in the choice of the charged isolation sideband and
the $\sigma_{\eta \eta}$ distribution used for estimating the fake photon probability.
The total uncertainties
in the fake photon background estimation can be found in Table~\ref{tab:sys_unc}.
The theoretical uncertainty in the top quark background is 20\%~\cite{Chatrchyan:2014bza}.

The systematic uncertainties in the estimation of  the trigger efficiency,
measured using the tag-and-probe technique, are 1.2\% and 1.7\%
for the $\Z\to\Pgmp\Pgmm$ and $\Z\to\Pep\Pem$ channels, respectively.
Using similar methods, the systematic uncertainties
in the efficiencies for lepton reconstruction and identification in the two channels are 1.9\% and 1.0\%, respectively.
The systematic uncertainty in the jet energy scale and resolution is estimated by varying the jet energy scale and resolution up and down within their \pt- and $\eta$-dependent uncertainties~\cite{Chatrchyan:2011ds}. The uncertainty is 14\% for $M_{\jj} >400$\GeV.
Another source of uncertainty is the modelling of the pileup. The inelastic cross section is
varied by $\pm$5\% in order to evaluate this contribution.
The uncertainty in the integrated luminosity is 2.6\% \cite{CMS:LUM13001}.

There are also three sources of theoretical uncertainties applied to the signal only.
The PDF uncertainty for the signal is estimated with the CT10~\cite{Guzzi:2011sv} PDF set, following the asymmetric Hessian method introduced in Refs.~\cite{Martin:2009iq,Lai:2010nw}.
The scale uncertainty is evaluated by varying the renormalization and factorization scales independently by a factor of two.
The magnitude of the interference between QCD and EW $\Z\gamma\jj$ processes is assigned as systematic uncertainties
in the two $M_{\jj}$ ranges.

All the systematic uncertainties described are applied to both the signal significance measurement and the aQGC search.
They are also propagated to the uncertainty in the measured fiducial cross section, with the exception of the theoretical uncertainty associated with the signal cross section.

All the uncertainties in our analysis are summarized in Table~\ref{tab:sys_unc}.

\begin{table*}[htb]
\centering
\topcaption{Summary of the major uncertainties.}
  \begin{tabular}{lc}
  \hline
  Source                        & Uncertainty \\
  \hline
  QCD $\Z\gamma$ + jets normalization  & 22\% ($400 <M_{\jj}<800$\GeV)    \\
                                 &  24\% ($M_{\jj}>800$\GeV)   \\[1.5ex]
  Fake photon from jet       & 15\% (20--30\GeV)  \\
  ($\pt^{\gamma}$ dependent)    & 22\% (30--50\GeV) \\
                                  & 49\% ($>$50\GeV) \\
  \hline
  Trigger efficiency              & 1.2\% ($\Z\to\Pgmp\Pgmm$), 1.7\% ($\Z\to\Pep\Pem$)       \\
  Lepton selection efficiency     & 1.9\% ($\Z\to\Pgmp\Pgmm$), 1.0\% ($\Z\to\Pep\Pem$)         \\
  Jet energy scale and resolution  & 14\% ($M_{\jj}>400$\GeV)   \\
  $\ttbar\gamma$ cross section  & 20\%~\cite{Chatrchyan:2014bza}    \\
  Pileup modeling                 & 1.0\%         \\
  \hline
  Renormalization/    & 9.0\% ($400 <M_{\jj}<800$\GeV), 12\% ($M_{\jj}>800$\GeV) (SM)    \\
  factorization scale (signal)                                              & 14\% (aQGC)   \\[1.5ex]
  PDF (signal)                                    & 4.2\% ($400 <M_{\jj}<800$\GeV), 2.4\% ($M_{\jj}>800$\GeV) (SM) \\
                                             &4.3\% (aQGC)  \\[1.5ex]
  Interference (signal)                        & 18\% ($400 <M_{\jj}<800$\GeV), 11\% ($M_{\jj}>800$\GeV) (SM)    \\
  \hline
  Luminosity                    & 2.6\%         \\
  \hline
  \end{tabular}
  \label{tab:sys_unc}
\end{table*}

\section{Measurement of the signal significance and fiducial cross section}
\label{sec:sigif}

As shown in Table~\ref{tab:selection}, in addition to the common selection, we apply three further requirements to isolate the EW signal:
$\abs{y_{\Z\gamma}-(y_{\mathrm{j}1}+y_{\mathrm{j}2})/2}<1.2$,
$\abs{\Delta\eta_{\jj}}>1.6$, and $\Delta\phi_{\Z\gamma,\jj}>2.0$ radians.
The selection requirements are chosen by optimizing the expected significance.
We apply the CL$_\mathrm{s}$ criterion described in Ref.~\cite{CLs1,Junk:1999kv} to assess the signal significance, based on the binned $M_{\jj}$ distribution, using only the two rightmost bins corresponding to $400 <M_{\jj}<800$\GeV and $M_{\jj}>800$\GeV.
We consider QCD $\Z\gamma\jj$ production and events without $\Z\gamma$ as background and EW $\Z\gamma\jj$ production as signal.

Table~\ref{tab:yields} summarizes the number of events predicted for each process with the number of events observed.
For EW $\Z\gamma\jj$ production, the observations are found to be compatible with expectations in the different channels.
By combining both channels, we find evidence for EW $\Z\gamma\jj$ production with an observed and expected significance of 3.0 and 2.1 standard deviations, respectively.
We determine the ratio of the observed signal to that expected from the SM for LO
EW $\Z\gamma\jj$ production as $\hat{\mu}$ = $1.5_{-0.6}^{+0.9}$
using a binned likelihood fit over the two ranges of the $M_{\jj}$ distribution.

Applying the same criteria, we can also measure the significance of the combined EW and QCD $\Z\gamma\jj$ process.
As shown in Table~\ref{tab:yields}, with the two decay channels combined in the search region,
of the signal events 7.0 (38.4\%) are estimated to come from EW production and the remaining 11.3 from QCD production.
As a result, the observed (expected) significance for the combined EW and QCD $\Z\gamma\jj$ process is 5.7 (5.5) standard deviations.

\begin{table}[htbp]
\begin{center}
\topcaption{Signal and background yields after the final selection for the SM measurement, for the two bins of 400 $<M_{\jj}<800$\GeV (upper) and $M_{\jj}>800$\GeV (lower). Only statistical uncertainties are reported.}
\begin{tabular}{lcc}
\hline
$400 <M_{\jj}<800$\GeV & muon & electron\\
\hline
Fake photon from jet      & 3.4 $\pm$ 0.8  & 1.7 $\pm$ 0.5\\
Other background           & 0.1 $\pm$ 0.1  & 0.1 $\pm$ 0.1\\
QCD $\Z\gamma\jj$            & 4.8 $\pm$ 0.9  & 5.0 $\pm$ 1.0\\
EW $\Z\gamma\jj$             & 1.7 $\pm$ 0.1  & 1.8 $\pm$ 0.1\\
Total background           & 8.3 $\pm$ 1.2  & 6.8 $\pm$ 1.1\\
Data                            & 13         & 8         \\
\hline
$M_{\jj}>800$\GeV & muon  & electron \\
\hline
Fake photon from jet      & 0.4 $\pm$ 0.3  & 0.1 $\pm$ 0.1\\
Other background           & 0 $\pm$ 0  & 0 $\pm$ 0\\
QCD $\Z\gamma\jj$            & 0.4 $\pm$ 0.1  & 1.1 $\pm$ 0.2 \\
EW $\Z\gamma\jj$             & 1.8 $\pm$ 0.1  & 1.8 $\pm$ 0.1\\
Total background           & 0.8 $\pm$ 0.3  & 1.2 $\pm$ 0.2\\
Data                            & 5                 & 2         \\
\hline
\end{tabular}
\label{tab:yields}
\end{center}
\end{table}

To determine the cross section for EW $\Z\gamma\jj$ production
we use a fiducial kinematic region based on the acceptance of the CMS detector with a minimal selection on the $M_{\jj}$ and $\Delta\eta_{\jj}$ variables to select the VBS topology. The fiducial region is defined as described in Table~\ref{tab:selection}.
We define the cross section in the fiducial region as $\sigma_{f} =\sigma_{g} \,  \hat{\mu} \, \alpha_{gf}$ where
$\sigma_{g}$ is the cross section for generated signal events,
$\hat{\mu}$ is the signal strength, and
$\alpha_{gf}$ is the acceptance for the generated events in the fiducial region, evaluated through simulation.
The fiducial cross section for EW $\Z\gamma\jj$ production is
$1.86_{-0.75}^{+0.90}\stat_{-0.26}^{+0.34}\syst \pm 0.05\lum$\unit{fb},
consistent with the theoretical prediction at LO of $1.27 \pm 0.11\,\text{(scale)}\pm 0.05\,\mathrm{(PDF)}$\unit{fb} calculated using \MADGRAPH.

The cross section for all processes that produce the $\Z\gamma\jj$ final state can be compared to theoretical predictions.
The fiducial region studied here lies in a particularly interesting region of phase space because of the substantial contribution to $\Z\gamma\jj$ from EW production.
By restricting the phase space to the fiducial region for the EW process as defined before, the expected fraction of EW events in the combined sample of EW and QCD signal events is 26\%, and the cross section of the combined process is
$5.94_{-1.35}^{+1.53}\stat_{-0.37}^{+0.43}\syst\pm 0.13\lum$\unit{fb},
which is consistent with the theoretical prediction at LO calculated using \MADGRAPH:
$5.05 \pm 1.22\,\text{(scale)}\pm 0.31\,\mathrm{(PDF)}$\unit{fb}.

\section{Search for anomalous quartic gauge couplings}
\label{sec:aqgcparam}

The effects of any new physics between the \TeVns and the Planck scale might be significant in the high energy tails of measurements at the LHC and can be parameterized via effective anomalous couplings.
With the discovery of the Higgs boson, higher-dimensional operators can be introduced in a linear way~\cite{Eboli:2006wa}:
\ifthenelse{\boolean{cms@external}}{
\begin{equation}\label{lagrangian}
\begin{aligned}
\mathcal{L}_{\mathrm{aQGC}} =&
 \frac{f_\mathrm{M0}}{\Lambda^{4}}\;\Tr\left[ \mathbf{W}_{\mu\nu} \mathbf{W}^{\mu\nu} \right] \times \left[  (D_{\beta}\Phi)^{\dagger} D^{\beta}\Phi \right] \\
 +& \frac{f_\mathrm{M1}}{\Lambda^{4}}\;\Tr\left[ \mathbf{W}_{\mu\nu} \mathbf{W}^{\nu\beta}\right] \times \left[  (D_{\beta}\Phi)^{\dagger} D^{\mu}\Phi \right] \ \\
 +& \frac{f_\mathrm{M2}}{\Lambda^{4}}\;\left[ B_{\mu\nu} B^{\mu\nu}\right] \times \left[  (D_{\beta}\Phi)^{\dagger} D^{\beta}\Phi \right] \\
 +& \frac{f_\mathrm{M3}}{\Lambda^{4}}\;\left[ B_{\mu\nu} B^{\nu\beta}\right] \times \left[  (D_{\beta}\Phi)^{\dagger} D^{\mu}\Phi \right] \ \\
 +& \frac{f_\mathrm{T0}}{\Lambda^4} Tr[\hat{W}_{\mu \nu} \hat{W}^{\mu \nu}] \times Tr[\hat{W}_{\alpha \beta} \hat{W}^{\alpha \beta}] \\
 +& \frac{f_\mathrm{T2}}{\Lambda^{4}}\; Tr[\hat{W}_{\alpha \mu} \hat{W}^{\mu \beta}] \times Tr[\hat{W}_{\beta \nu} \hat{W}^{\nu \alpha}] \ \\
 +& \frac{f_\mathrm{T8}}{\Lambda^4} B_{\mu\nu}B^{\mu\nu}B_{\alpha\beta}B^{\alpha\beta} + \frac{f_\mathrm{T9}}{\Lambda^4} B_{\alpha\mu}B^{\mu\beta}B_{\beta\nu}B^{\nu\alpha},
\end{aligned}
\end{equation}
}{
\begin{equation}\label{lagrangian}
\begin{aligned}
\mathcal{L}_{\mathrm{aQGC}} =&
 \frac{f_\mathrm{M0}}{\Lambda^{4}}\;\Tr\left[ \mathbf{W}_{\mu\nu} \mathbf{W}^{\mu\nu} \right] \times \left[  (D_{\beta}\Phi)^{\dagger} D^{\beta}\Phi \right] +
 \frac{f_\mathrm{M1}}{\Lambda^{4}}\;\Tr\left[ \mathbf{W}_{\mu\nu} \mathbf{W}^{\nu\beta}\right] \times \left[  (D_{\beta}\Phi)^{\dagger} D^{\mu}\Phi \right] \ \\
 +& \frac{f_\mathrm{M2}}{\Lambda^{4}}\;\left[ B_{\mu\nu} B^{\mu\nu}\right] \times \left[  (D_{\beta}\Phi)^{\dagger} D^{\beta}\Phi \right] +
 \frac{f_\mathrm{M3}}{\Lambda^{4}}\;\left[ B_{\mu\nu} B^{\nu\beta}\right] \times \left[  (D_{\beta}\Phi)^{\dagger} D^{\mu}\Phi \right] \ \\
 +& \frac{f_\mathrm{T0}}{\Lambda^4} Tr[\hat{W}_{\mu \nu} \hat{W}^{\mu \nu}] \times Tr[\hat{W}_{\alpha \beta} \hat{W}^{\alpha \beta}] + \frac{f_\mathrm{T2}}{\Lambda^{4}}\; Tr[\hat{W}_{\alpha \mu} \hat{W}^{\mu \beta}] \times Tr[\hat{W}_{\beta \nu} \hat{W}^{\nu \alpha}] \ \\
 +& \frac{f_\mathrm{T8}}{\Lambda^4} B_{\mu\nu}B^{\mu\nu}B_{\alpha\beta}B^{\alpha\beta} + \frac{f_\mathrm{T9}}{\Lambda^4} B_{\alpha\mu}B^{\mu\beta}B_{\beta\nu}B^{\nu\alpha},
\end{aligned}
\end{equation}
}
where $f_\mathrm{M0,1,2,3}$ and $f_\mathrm{T0,2,8,9}$ are coefficients of relevant effective operators, and $\Lambda$ represents the scale of new physics responsible for anomalous couplings.
The Lagrangian of the aQGCs is implemented within the \MADGRAPH package.

\begin{figure}[htb]
  \begin{center}
	  \includegraphics[width=\cmsFigWidth]{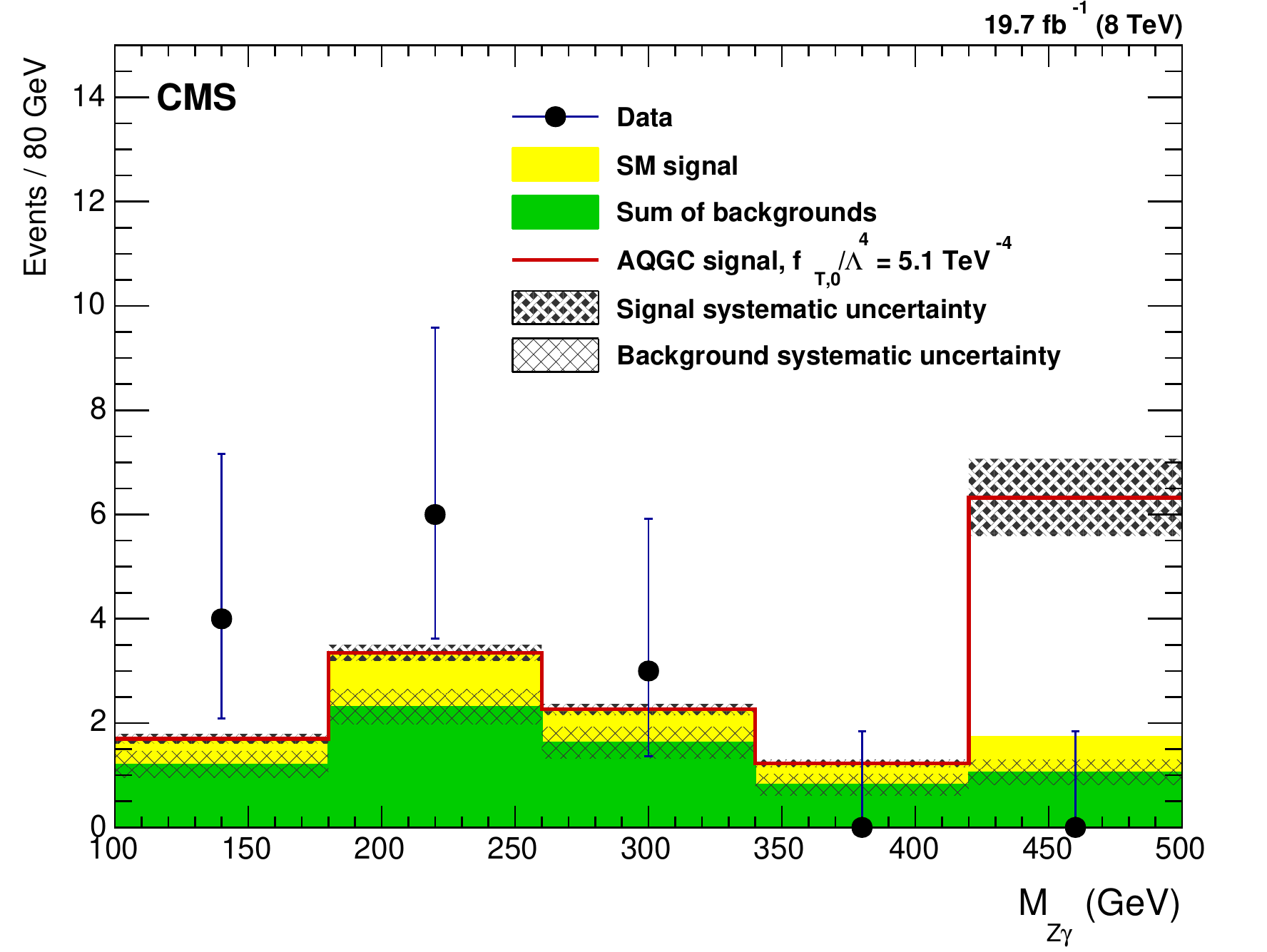} \\
    \caption{
The invariant mass distribution of the $\Z\gamma$ system for events that pass the aQGC selection.
The highest mass bin includes events with $M_{\Z\gamma} > 420$\GeV.
Error bars represent the statistical uncertainty in the data, while the systematic uncertainties in the aQGC signal and background estimate are shown as hatched bands. }
		\label{fig:zam_shape}
  \end{center}
\end{figure}

We study the distribution of the mass of the dilepton and photon system, $M_{\Z\gamma}$, to search for contributions from aQGCs.
The effects of new physics would be seen at higher energy and modify the interference of VBS diagrams. To select the region sensitive to new physics, we require $\pt^{\gamma} > 60$\GeV.
The selection for the aQGC analysis is described in Table~\ref{tab:selection}.
The $\Z\gamma$ mass distribution is shown in Fig. \ref{fig:zam_shape}, where the last bin includes all events with $M_{\Z\gamma}>420$\GeV.
Because no significant excess is seen in the $M_{\Z\gamma}$ distribution, we use the shape of the $M_{\Z\gamma}$ distribution to extract limits on aQGC contributions.

With the parameterization of signals and related systematic uncertainties,
for each aQGC parameter, we reweight the SM signal shape to the aQGC shape.
The following test statistic is used:
\begin{equation}
t_{\alpha_{\text{test}}} = -2 \ln\frac{\mathcal{L}(\alpha_{\text{test}},{\hat{\hat{\boldsymbol{\theta}}}})}{\mathcal{L}(\hat{\alpha},\hat{\boldsymbol{\theta}})},
\end{equation}
where the likelihood function ($\mathcal{L}$) is constructed for both lepton channels and combined,
using a bin-wise Poisson distribution with profiled nuisance parameters ($\boldsymbol{\theta}$).
$\alpha_{\text{test}}$ represents the aQGC point being tested.
The symbol $\hat{\hat{\boldsymbol{\theta}}}$ represents the values corresponding to the maximum of the likelihood at the point $\alpha_{\text{test}}$, while $\hat{\alpha}$ and $\hat{\boldsymbol{\theta}}$ correspond to the global maximum of the likelihood.
This test statistic is assumed to follow a $\chi^2$ distribution~\cite{1943Wald:wilks1938}, from which one can extract limits.
Exclusion limits
are shown in
Table~\ref{tab:limit_noff}.
Each coupling parameter is
varied over a set of discrete values, keeping the other parameters fixed to zero.

\begin{table}[htb]
\centering
  \topcaption{Observed and expected shape-based exclusion limits for each aQGC parameter at 95\% CL, without a form factor applied.}
  \begin{tabular}{cc}
	 \hline
	 Observed limits ($\TeVns^{-4}$) & Expected limits ($\TeVns^{-4}$) \\
	 \hline
	$-71<f_\mathrm{M0}/\Lambda^{4}<75$ & $-109<f_\mathrm{M0}/\Lambda^{4}<111$   \\
$-190<f_\mathrm{M1}/\Lambda^{4}<182$  &  $-281<f_\mathrm{M1}/\Lambda^{4}<280$    \\
 $-32<f_\mathrm{M2}/\Lambda^{4}<31$   &   $-47<f_\mathrm{M2}/\Lambda^{4}<47$   \\
 $-58<f_\mathrm{M3}/\Lambda^{4}<59$  & $-87<f_\mathrm{M3}/\Lambda^{4}<87$   \\
 $-3.8<f_\mathrm{T0}/\Lambda^{4}<3.4$  & $-5.1<f_\mathrm{T0}/\Lambda^{4}<5.1$    \\
 $-4.4<f_\mathrm{T1}/\Lambda^{4}<4.4$  & $-6.5<f_\mathrm{T1}/\Lambda^{4}<6.5$    \\
 $-9.9<f_\mathrm{T2}/\Lambda^{4}<9.0$  & $-14.0<f_\mathrm{T2}/\Lambda^{4}<14.5$    \\
 $-1.8<f_\mathrm{T8}/\Lambda^{4}<1.8$  &  $-2.7<f_\mathrm{T8}/\Lambda^{4}<2.7$    \\
 $-4.0<f_\mathrm{T9}/\Lambda^{4}<4.0$  &  $-6.0<f_\mathrm{T9}/\Lambda^{4}<6.0$    \\	
\hline
\end{tabular}
 \label{tab:limit_noff}
\end{table}

An effective theory is only valid at energies lower than the scale of new physics,
and high-dimensional operators with nonzero aQGC values can lead to unitarity violation at sufficiently high energies.
For each aQGC listed in Table~\ref{tab:limit_noff}, we checked the stated upper limit against the unitary bound~\cite{Gounaris:1993fh} obtained with \textsc{vbfnlo}~\cite{Baglio:2014uba}.
In general, we find the limits on all aQGC parameters are set in the unitary unsafe region, except for $f_\mathrm{T9}$ where the unitarity bound is up to 6\TeV.
Form factors can be introduced to unitarize the high energy contribution,
however it is difficult to compare results from different experiments and it is not theoretically well motivated. In this study all of the aQGC limits shown are evaluated without a form factor, and can be directly compared to limits set in references~\cite{Chatrchyan:2014bza,Aad:2015uqa,Aad:2016sau,CMS-PAS-SMP-15-008,Aad:2014zda,CMS-PAS-SMP-13-015,Khachatryan:2016mud,CMS-PAS-SMP-14-011,Aad:2016ett}.

\section{Conclusions}
\label{sec:sum}

The measurement of the cross section for the electroweak production
of a Z boson and a photon in association with two jets,
where the \Z boson decays into electron or muon pairs, was presented.
The measurement is based on a sample of
proton-proton collisions collected with the CMS detector at a center-of-mass energy of 8\TeV, corresponding to an integrated luminosity of
19.7\fbinv.
We find evidence for EW $\Z\gamma\jj$ production with an observed (expected) significance of 3.0 (2.1) standard deviations.
The fiducial cross section for EW $\Z\gamma\jj$ production is measured to be $1.86_{-0.75}^{+0.90}\stat_{-0.26}^{+0.34}\syst \pm 0.05\lum$\unit{fb}, consistent with the theoretical prediction.
The fiducial cross section for combined EW and QCD $\Z\gamma\jj$ production is $5.94_{-1.35}^{+1.53}\stat_{-0.37}^{+0.43}\syst \pm 0.13\lum$\unit{fb}, which is also consistent with the leading-order theoretical prediction.

In the framework of dimension-eight effective field theory operators, limits on the aQGC parameters $f_\mathrm{M0,1,2,3}$ and $f_\mathrm{T0,1,2,8,9}$ are set at 95\% confidence level.
This is the first constraints on the neutral aQGC parameters $f_\mathrm{T8}$.

\begin{acknowledgments}
We congratulate our colleagues in the CERN accelerator departments for the excellent performance of the LHC and thank the technical and administrative staffs at CERN and at other CMS institutes for their contributions to the success of the CMS effort. In addition, we gratefully acknowledge the computing centers and personnel of the Worldwide LHC Computing Grid for delivering so effectively the computing infrastructure essential to our analyses. Finally, we acknowledge the enduring support for the construction and operation of the LHC and the CMS detector provided by the following funding agencies: BMWFW and FWF (Austria); FNRS and FWO (Belgium); CNPq, CAPES, FAPERJ, and FAPESP (Brazil); MES (Bulgaria); CERN; CAS, MoST, and NSFC (China); COLCIENCIAS (Colombia); MSES and CSF (Croatia); RPF (Cyprus); SENESCYT (Ecuador); MoER, ERC IUT, and ERDF (Estonia); Academy of Finland, MEC, and HIP (Finland); CEA and CNRS/IN2P3 (France); BMBF, DFG, and HGF (Germany); GSRT (Greece); OTKA and NIH (Hungary); DAE and DST (India); IPM (Iran); SFI (Ireland); INFN (Italy); MSIP and NRF (Republic of Korea); LAS (Lithuania); MOE and UM (Malaysia); BUAP, CINVESTAV, CONACYT, LNS, SEP, and UASLP-FAI (Mexico); MBIE (New Zealand); PAEC (Pakistan); MSHE and NSC (Poland); FCT (Portugal); JINR (Dubna); MON, RosAtom, RAS, RFBR and RAEP (Russia); MESTD (Serbia); SEIDI, CPAN, PCTI and FEDER (Spain); Swiss Funding Agencies (Switzerland); MST (Taipei); ThEPCenter, IPST, STAR, and NSTDA (Thailand); TUBITAK and TAEK (Turkey); NASU and SFFR (Ukraine); STFC (United Kingdom); DOE and NSF (USA).

\hyphenation{Rachada-pisek} Individuals have received support from the Marie-Curie program and the European Research Council and EPLANET (European Union); the Leventis Foundation; the A. P. Sloan Foundation; the Alexander von Humboldt Foundation; the Belgian Federal Science Policy Office; the Fonds pour la Formation \`a la Recherche dans l'Industrie et dans l'Agriculture (FRIA-Belgium); the Agentschap voor Innovatie door Wetenschap en Technologie (IWT-Belgium); the Ministry of Education, Youth and Sports (MEYS) of the Czech Republic; the Council of Science and Industrial Research, India; the HOMING PLUS program of the Foundation for Polish Science, cofinanced from European Union, Regional Development Fund, the Mobility Plus program of the Ministry of Science and Higher Education, the National Science Center (Poland), contracts Harmonia 2014/14/M/ST2/00428, Opus 2014/13/B/ST2/02543, 2014/15/B/ST2/03998, and 2015/19/B/ST2/02861, Sonata-bis 2012/07/E/ST2/01406; the National Priorities Research Program by Qatar National Research Fund; the Programa Clar\'in-COFUND del Principado de Asturias; the Thalis and Aristeia programs cofinanced by EU-ESF and the Greek NSRF; the Rachadapisek Sompot Fund for Postdoctoral Fellowship, Chulalongkorn University and the Chulalongkorn Academic into Its 2nd Century Project Advancement Project (Thailand); and the Welch Foundation, contract C-1845.
\end{acknowledgments}

\bibliography{auto_generated}

\providecommand{\href}[2]{#2}\begingroup\raggedright\begin{thebibliography}{10}%
\makeatletter
\providecommand{\hrefCMSnoop }[0]{\@secondoftwo}%
\makeatother
\providecommand{\doi}{\texttt{doi:}\begingroup \urlstyle{tt}\Url}

\bibitem{Aad:2012tfa}
\hrefCMSnoop {}{{ATLAS Collaboration}, ``{Observation of a new particle in the
  search for the Standard Model Higgs boson with the ATLAS detector at the
  LHC}'',} \textit{ Phys. Lett. B} \textbf{ 716} (2012) 1,
  \href{http://dx.doi.org/10.1016/j.physletb.2012.08.020}{\doi{10.1016/j.physletb.2012.08.020}},
\href{http://www.arXiv.org/abs/1207.7214}{\texttt{arXiv:1207.7214}}.
%%CITATION = ARXIV:1207.7214;%%.

\bibitem{Chatrchyan:2012ufa}
\hrefCMSnoop {}{{CMS Collaboration}, ``{Observation of a new boson at a mass of
  125 GeV with the CMS experiment at the LHC}'',} \textit{ Phys. Lett. B}
  \textbf{ 716} (2012) 30,
  \href{http://dx.doi.org/10.1016/j.physletb.2012.08.021}{\doi{10.1016/j.physletb.2012.08.021}},
\href{http://www.arXiv.org/abs/1207.7235}{\texttt{arXiv:1207.7235}}.
%%CITATION = ARXIV:1207.7235;%%.

\bibitem{Chatrchyan:2014bza}
\hrefCMSnoop {}{{CMS Collaboration}, ``{A search for $WW\gamma$ and $WZ\gamma$
  production and constraints on anomalous quartic gauge couplings in pp
  collisions at $\sqrt{s} =8$ TeV}'',} \textit{ Phys. Rev. D} \textbf{ 90}
  (2014) 032008,
  \href{http://dx.doi.org/10.1103/PhysRevD.90.032008}{\doi{10.1103/PhysRevD.90.032008}},
\href{http://www.arXiv.org/abs/1404.4619}{\texttt{arXiv:1404.4619}}.
%%CITATION = ARXIV:1404.4619;%%.

\bibitem{Aad:2015uqa}
\hrefCMSnoop {}{{ATLAS Collaboration}, ``Evidence of {$W\gamma\gamma$}
  production in pp collisions at {$\sqrt{s} = 8~\mathrm{TeV}$} and limits on
  anomalous quartic gauge couplings with the {ATLAS} detector'',} \textit{
  Phys. Rev. Lett.} \textbf{ 115} (2015) 031802,
  \href{http://dx.doi.org/10.1103/PhysRevLett.115.031802}{\doi{10.1103/PhysRevLett.115.031802}},
\href{http://www.arXiv.org/abs/1503.03243}{\texttt{arXiv:1503.03243}}.
%%CITATION = ARXIV:1503.03243;%%.

\bibitem{Aad:2016sau}
\hrefCMSnoop {}{{ATLAS Collaboration}, ``{Measurements of $Z\gamma$ and
  $Z\gamma\gamma$ production in pp collisions at $\sqrt{s}=$ 8 TeV with the
  ATLAS detector}'',} \textit{ Phys. Rev. D} \textbf{ 93} (2016) 112002,
  \href{http://dx.doi.org/10.1103/PhysRevD.93.112002}{\doi{10.1103/PhysRevD.93.112002}},
\href{http://www.arXiv.org/abs/1604.05232}{\texttt{arXiv:1604.05232}}.
%%CITATION = doi:10.1103/PhysRevD.93.112002;%%.

\bibitem{CMS-PAS-SMP-15-008}
\href {http://cds.cern.ch/record/2130360}{{CMS Collaboration}, ``Measurements
  of the {pp$\to\mathrm{W}^{\pm}\gamma\gamma$} and
  {pp$\to\mathrm{Z}\gamma\gamma$} cross sections and limits on dimension-8
  effective anomalous gauge couplings at {$\sqrt{s} = 8~\mathrm{TeV}$}'',} CMS
  Physics Analysis Summary CMS-PAS-SMP-15-008, 2016.

\bibitem{Aad:2014zda}
\hrefCMSnoop {}{{ATLAS Collaboration}, ``Evidence for electroweak production of
  {$W^{\pm}W^{\pm}$jj} in $pp$ collisions at {$\sqrt{s} = 8$ TeV} with the
  {ATLAS} detector'',} \textit{ Phys. Rev. Lett.} \textbf{ 113} (2014) 141803,
  \href{http://dx.doi.org/10.1103/PhysRevLett.113.141803}{\doi{10.1103/PhysRevLett.113.141803}},
\href{http://www.arXiv.org/abs/1405.6241}{\texttt{arXiv:1405.6241}}.
%%CITATION = ARXIV:1405.6241;%%.

\bibitem{Aaboud:2016ffv}
\hrefCMSnoop {}{{ATLAS Collaboration}, ``{Measurement of $W^{\pm}W^{\pm}$
  vector-boson scattering and limits on anomalous quartic gauge couplings with
  the ATLAS detector}'',}
\href{http://www.arXiv.org/abs/1611.02428}{\texttt{arXiv:1611.02428}}.
%%CITATION = ARXIV:1611.02428;%%.

\bibitem{CMS-PAS-SMP-13-015}
\hrefCMSnoop {}{{CMS Collaboration}, ``Study of vector boson scattering and
  search for new physics in events with two same-sign leptons and two jets'',}
  \textit{ Phys. Rev. Lett.} \textbf{ 114} (2014) 051801,
  \href{http://dx.doi.org/10.1103/PhysRevLett.114.051801}{\doi{10.1103/PhysRevLett.114.051801}},
\href{http://www.arXiv.org/abs/1410.6315}{\texttt{arXiv:1410.6315}}.
%%CITATION = ARXIV:1410.6315;%%.

\bibitem{Khachatryan:2016mud}
\hrefCMSnoop {}{{CMS Collaboration}, ``{Evidence for exclusive $\gamma\gamma$
  to $W^{+}W^{-}$ production and constraints on anomalous quartic gauge
  couplings at $\sqrt{s} = 7$ and 8 TeV}'',} \textit{ JHEP} \textbf{ 08} (2016)
  119,
  \href{http://dx.doi.org/10.1007/JHEP08(2016)119}{\doi{10.1007/JHEP08(2016)119}},
\href{http://www.arXiv.org/abs/1604.04464}{\texttt{arXiv:1604.04464}}.
%%CITATION = doi:10.1007/JHEP08(2016)119;%%.

\bibitem{Khachatryan:2016vif}
\hrefCMSnoop {}{{CMS Collaboration}, ``{Measurement of electroweak-induced
  production of W gamma with two jets in pp collisions at $\sqrt{s}$ = 8 TeV
  and constraints on anomalous quartic gauge couplings}'',}
\href{http://www.arXiv.org/abs/1612.09256}{\texttt{arXiv:1612.09256}}.
%%CITATION = ARXIV:1612.09256;%%.

\bibitem{Aad:2016ett}
\hrefCMSnoop {}{{ATLAS Collaboration}, ``{Measurements of $W^\pm Z$ production
  cross sections in $pp$ collisions at $\sqrt{s} = 8$ TeV with the ATLAS
  detector and limits on anomalous gauge boson self-couplings}'',} \textit{
  Phys. Rev. D} \textbf{ 93} (2016) 092004,
  \href{http://dx.doi.org/10.1103/PhysRevD.93.092004}{\doi{10.1103/PhysRevD.93.092004}},
\href{http://www.arXiv.org/abs/1603.02151}{\texttt{arXiv:1603.02151}}.
%%CITATION = ARXIV:1603.02151;%%.

\bibitem{Chatrchyan:2013jya}
\hrefCMSnoop {}{{CMS Collaboration}, ``{Measurement of the hadronic activity in
  events with a Z and two jets and extraction of the cross section for the
  electroweak production of a Z with two jets in pp collisions at $\sqrt{s} =
  7$ TeV}'',} \textit{ JHEP} \textbf{ 10} (2013) 062,
  \href{http://dx.doi.org/10.1007/JHEP10(2013)062}{\doi{10.1007/JHEP10(2013)062}},
\href{http://www.arXiv.org/abs/1305.7389}{\texttt{arXiv:1305.7389}}.
%%CITATION = ARXIV:1305.7389;%%.

\bibitem{CMS-PAS-FSQ-12-035}
\hrefCMSnoop {}{{CMS Collaboration}, ``Measurement of electroweak production of
  two jets in association with a {Z} boson in proton-proton collisions at
  {$\sqrt{s} = 8$ TeV}'',} \textit{ Eur. Phys. J. C} \textbf{ 75} (2014) 2,
  \href{http://dx.doi.org/10.1140/epjc/s10052-014-3232-5}{\doi{10.1140/epjc/s10052-014-3232-5}},
\href{http://www.arXiv.org/abs/1410.3153}{\texttt{arXiv:1410.3153}}.
%%CITATION = ARXIV: 1410.3153;%%.

\bibitem{Aad:2014dta}
\hrefCMSnoop {}{{ATLAS Collaboration}, ``Measurement of the electroweak
  production of dijets in association with a {$Z$}-boson and distributions
  sensitive to vector boson fusion in proton-proton collisions at {$\sqrt{s} =
  8$ TeV} using the {ATLAS} detector'',} \textit{ JHEP} \textbf{ 04} (2013)
  031,
  \href{http://dx.doi.org/10.1007/JHEP04(2014)031}{\doi{10.1007/JHEP04(2014)031}},
\href{http://www.arXiv.org/abs/1401.7610}{\texttt{arXiv:1401.7610}}.
%%CITATION = ARXIV:1401.7610;%%.

\bibitem{CMS-PAS-SMP-13-012}
\hrefCMSnoop {}{{CMS Collaboration}, ``{Measurement of electroweak production
  of a W boson and two forward jets in proton-proton collisions at $ \sqrt{s}=8
  $ TeV}'',} \textit{ JHEP} \textbf{ 11} (2016) 147,
  \href{http://dx.doi.org/10.1007/JHEP11(2016)147}{\doi{10.1007/JHEP11(2016)147}},
\href{http://www.arXiv.org/abs/1607.06975}{\texttt{arXiv:1607.06975}}.
%%CITATION = ARXIV:1607.06975;%%.

\bibitem{CMS-PAS-PFT-09-001}
\href {http://cds.cern.ch/record/1194487}{{CMS Collaboration}, ``Particle--flow
  event reconstruction in {CMS} and performance for jets, taus, and {\MET}'',}
  CMS Physics Analysis Summary CMS-PAS-PFT-09-001, 2009.

\bibitem{CMS-PAS-PFT-10-001}
\href {http://cds.cern.ch/record/1247373}{{CMS Collaboration}, ``Commissioning
  of the particle-flow event reconstruction with the first {LHC} collisions
  recorded in the {CMS} detector'',} CMS Physics Analysis Summary
  CMS-PAS-PFT-10-001, 2010.

\bibitem{CMS:EGM-14-001}
\hrefCMSnoop {}{{CMS Collaboration}, ``{Performance of photon reconstruction
  and identification with the CMS detector in proton-proton collisions at
  $\sqrt{s} = 8$\TeV}'',} \textit{ JINST} \textbf{ 10} (2015) P08010,
  \href{http://dx.doi.org/10.1088/1748-0221/10/08/P08010}{\doi{10.1088/1748-0221/10/08/P08010}},
\href{http://www.arXiv.org/abs/1502.02702}{\texttt{arXiv:1502.02702}}.
%%CITATION = ARXIV:1502.02702;%%.

\bibitem{Chatrchyan:2008zzk}
\hrefCMSnoop {}{{CMS Collaboration}, ``The {CMS} experiment at the {CERN}
  {LHC}'',} \textit{ JINST} \textbf{ 3} (2008) S08004,
\href{http://dx.doi.org/10.1088/1748-0221/3/08/S08004}{\doi{10.1088/1748-0221/3/08/S08004}}.
%%CITATION = JINST,3,S08004;%%.

\bibitem{Chatrchyan:2012xi}
\hrefCMSnoop {}{{CMS Collaboration}, ``{Performance of CMS muon reconstruction
  in pp collision events at $\sqrt{s} = 7$ TeV}'',} \textit{ JINST} \textbf{ 7}
  (2012) P10002,
  \href{http://dx.doi.org/10.1088/1748-0221/7/10/P10002}{\doi{10.1088/1748-0221/7/10/P10002}},
\href{http://www.arXiv.org/abs/1206.4071}{\texttt{arXiv:1206.4071}}.
%%CITATION = ARXIV:1206.4071;%%.

\bibitem{Chatrchyan:2015jya}
\hrefCMSnoop {}{{CMS Collaboration}, ``Measurement of the {Z$\gamma$}
  production cross section in pp collisions at {8 TeV} and search for anomalous
  triple gauge boson couplings'',} \textit{ JHEP} \textbf{ 04} (2015) 164,
  \href{http://dx.doi.org/10.1007/JHEP04(2015)164}{\doi{10.1007/JHEP04(2015)164}},
\href{http://www.arXiv.org/abs/1502.05664}{\texttt{arXiv:1502.05664}}.
%%CITATION = ARXIV: 1502.05664;%%.

\bibitem{Chatrchyan:2013dga}
\hrefCMSnoop {}{{CMS Collaboration}, ``{Energy calibration and resolution of
  the CMS electromagnetic calorimeter in pp collision at $\sqrt{s}= 7$ TeV}'',}
  \textit{ JINST} \textbf{ 8} (2013) P09009,
  \href{http://dx.doi.org/10.1088/1748-0221/8/09/P09009}{\doi{10.1088/1748-0221/8/09/P09009}},
\href{http://www.arXiv.org/abs/1306.2016}{\texttt{arXiv:1306.2016}}.
%%CITATION = ARXIV:1306.2016;%%.

\bibitem{VBTF}
\hrefCMSnoop {}{{CMS Collaboration}, ``{Measurement of the Inclusive W and Z
  Production Cross Sections in pp Collisions at $\sqrt{s}$ = 7 TeV}'',}
  \textit{ JHEP} \textbf{ 10} (2011) 132,
  \href{http://dx.doi.org/10.1007/JHEP10(2011)132}{\doi{10.1007/JHEP10(2011)132}},
\href{http://www.arXiv.org/abs/1107.4789}{\texttt{arXiv:1107.4789}}.
%%CITATION = ARXIV:1107.4789;%%.

\bibitem{Khachatryan:2015hwa}
\hrefCMSnoop {}{{CMS Collaboration}, ``Performance of electron reconstruction
  and selection with the {CMS} detector in proton-proton collisions at
  {$\sqrt{s} = 8$ TeV}'',} \textit{ JINST} \textbf{ 10} (2015) P06005,
  \href{http://dx.doi.org/10.1088/1748-0221/10/06/P06005}{\doi{10.1088/1748-0221/10/06/P06005}},
\href{http://www.arXiv.org/abs/1502.02701}{\texttt{arXiv:1502.02701}}.
%%CITATION = ARXIV:1502.02701;%%.

\bibitem{Cacciari:2011ma}
\hrefCMSnoop {}{M.~Cacciari, G.~P. Salam, and G.~Soyez, ``{FastJet} user
  manual'',} \textit{ Eur. Phys. J. C} \textbf{ 72} (2012) 1896,
  \href{http://dx.doi.org/10.1140/epjc/s10052-012-1896-2}{\doi{10.1140/epjc/s10052-012-1896-2}},
\href{http://www.arXiv.org/abs/1111.6097}{\texttt{arXiv:1111.6097}}.
%%CITATION = ARXIV:1111.6097;%%.

\bibitem{Cacciari:2008gp}
\hrefCMSnoop {}{M.~Cacciari, G.~P. Salam, and G.~Soyez, ``{The anti-$\kt$ jet
  clustering algorithm}'',} \textit{ JHEP} \textbf{ 04} (2008) 063,
  \href{http://dx.doi.org/10.1088/1126-6708/2008/04/063}{\doi{10.1088/1126-6708/2008/04/063}},
\href{http://www.arXiv.org/abs/0802.1189}{\texttt{arXiv:0802.1189}}.
%%CITATION = ARXIV:0802.1189;%%.

\bibitem{Chatrchyan:2011ds}
\hrefCMSnoop {}{{CMS Collaboration}, ``Determination of jet energy calibration
  and transverse momentum resolution in {CMS}'',} \textit{ JINST} \textbf{ 6}
  (2011) P11002,
  \href{http://dx.doi.org/10.1088/1748-0221/6/11/P11002}{\doi{10.1088/1748-0221/6/11/P11002}},
\href{http://www.arXiv.org/abs/1107.4277}{\texttt{arXiv:1107.4277}}.
%%CITATION = 1107.4277;%%.

\bibitem{CMS-PAS-JME-10-003}
\href {http://cdsweb.cern.ch/record/1279362}{{CMS Collaboration}, ``Jet
  performance in pp collisions at $\sqrt{s}$ = 7 tev'',} CMS Physics Analysis
  Summary CMS-PAS-JME-10-003, 2010.

\bibitem{Campanario:2014wga}
\hrefCMSnoop {}{F.~Campanario, M.~Kerner, L.~D. Ninh, and D.~Zeppenfeld,
  ``{$Z\gamma$} production in association with two jets at next-to-leading
  order {QCD}'',} \textit{ Eur. Phys. J. C} \textbf{ 74} (2014), no.~9, 3085,
  \href{http://dx.doi.org/10.1140/epjc/s10052-014-3085-y}{\doi{10.1140/epjc/s10052-014-3085-y}},
\href{http://www.arXiv.org/abs/1407.7857}{\texttt{arXiv:1407.7857}}.
%%CITATION = ARXIV:1407.7857;%%.

\bibitem{mlm}
\hrefCMSnoop {}{M.~L. Mangano, M.~Moretti, F.~Piccinini, and M.~Treccani,
  ``{Matching matrix elements and shower evolution for top-quark production in
  hadronic collisions}'',} \textit{ JHEP} \textbf{ 01} (2007) 013,
  \href{http://dx.doi.org/10.1088/1126-6708/2007/01/013}{\doi{10.1088/1126-6708/2007/01/013}},
\href{http://www.arXiv.org/abs/hep-ph/0611129}{\texttt{arXiv:hep-ph/0611129}}.
%%CITATION = HEP-PH/0611129;%%.

\bibitem{Alwall:2007fs}
J.~Alwall\hrefCMSnoop {}{ {et~al.}, ``{Comparative study of various algorithms
  for the merging of parton showers and matrix elements in hadronic
  collisions}'',} \textit{ Eur. Phys. J. C} \textbf{ 53} (2008) 473,
  \href{http://dx.doi.org/10.1140/epjc/s10052-007-0490-5}{\doi{10.1140/epjc/s10052-007-0490-5}},
\href{http://www.arXiv.org/abs/0706.2569}{\texttt{arXiv:0706.2569}}.
%%CITATION = 0706.2569;%%.

\bibitem{Alwall:2011uj}
J.~Alwall\hrefCMSnoop {}{ {et~al.}, ``{MadGraph} 5: going beyond'',} \textit{
  JHEP} \textbf{ 06} (2011) 128,
  \href{http://dx.doi.org/10.1007/JHEP06(2011)128}{\doi{10.1007/JHEP06(2011)128}},
\href{http://www.arXiv.org/abs/1106.0522}{\texttt{arXiv:1106.0522}}.
%%CITATION = ARXIV:1106.0522;%%.

\bibitem{Sjostrand:2006za}
\hrefCMSnoop {}{T.~Sj{\"o}strand, S.~Mrenna, and P.~Skands, ``{PYTHIA} 6.4
  physics and manual'',} \textit{ JHEP} \textbf{ 05} (2006) 026,
  \href{http://dx.doi.org/10.1088/1126-6708/2006/05/026}{\doi{10.1088/1126-6708/2006/05/026}},
\href{http://www.arXiv.org/abs/hep-ph/0603175}{\texttt{arXiv:hep-ph/0603175}}.
%%CITATION = HEP-PH/0603175;%%.

\bibitem{Pumplin:2002vw}
J.~Pumplin\hrefCMSnoop {}{ {et~al.}, ``{New generation of parton distributions
  with uncertainties from global QCD analysis}'',} \textit{ JHEP} \textbf{ 07}
  (2002) 012,
  \href{http://dx.doi.org/10.1088/1126-6708/2002/07/012}{\doi{10.1088/1126-6708/2002/07/012}},
  \href{http://www.arXiv.org/abs/hep-ph/0201195}{\texttt{arXiv:hep-ph/0201195}}.

\bibitem{geant4}
\hrefCMSnoop {}{{GEANT4} Collaboration, ``{GEANT4} --- a simulation toolkit'',}
  \textit{ Nucl. Instrum. Meth. A} \textbf{ 506} (2003) 250,
\href{http://dx.doi.org/10.1016/S0168-9002(03)01368-8}{\doi{10.1016/S0168-9002(03)01368-8}}.
%%CITATION = NUIMA,A506,250;%%.

\bibitem{Vgamma}
\hrefCMSnoop {}{{CMS Collaboration}, ``{Measurement of $W\gamma$ and $Z\gamma$
  production in pp collisions at $\sqrt{s} = 7$ TeV}'',} \textit{ Phys. Lett.
  B} \textbf{ 701} (2011) 535,
  \href{http://dx.doi.org/10.1016/j.physletb.2011.06.034}{\doi{10.1016/j.physletb.2011.06.034}},
\href{http://www.arXiv.org/abs/1105.2758}{\texttt{arXiv:1105.2758}}.
%%CITATION = ARXIV:1105.2758;%%.

\bibitem{CMS:LUM13001}
\href {http://cds.cern.ch/record/1598864}{{CMS Collaboration}, ``{CMS
  Luminosity Based on Pixel Cluster Counting --- Summer 2013 Update}'',} CMS
  Physics Analysis Summary CMS-PAS-LUM-13-001, 2013.

\bibitem{Guzzi:2011sv}
M.~Guzzi\hrefCMSnoop {}{ {et~al.}, ``{CT10} parton distributions and other
  developments in the global {QCD} analysis'',} (2011).
  \href{http://www.arXiv.org/abs/1101.0561}{\texttt{arXiv:1101.0561}}.

\bibitem{Martin:2009iq}
\hrefCMSnoop {}{A.~D. Martin, W.~J. Stirling, R.~S. Thorne, and G.~Watt,
  ``Parton distributions for the {LHC}'',} \textit{ Eur. Phys. J. C} \textbf{
  63} (2009) 189,
  \href{http://dx.doi.org/10.1140/epjc/s10052-009-1072-5}{\doi{10.1140/epjc/s10052-009-1072-5}},
\href{http://www.arXiv.org/abs/0901.0002}{\texttt{arXiv:0901.0002}}.
%%CITATION = ARXIV:0901.0002;%%.

\bibitem{Lai:2010nw}
H.-L. Lai\hrefCMSnoop {}{ {et~al.}, ``Uncertainty induced by {QCD} coupling in
  the {CTEQ} global analysis of parton distributions'',} \textit{ Phys. Rev. D}
  \textbf{ 82} (2010) 054021,
  \href{http://dx.doi.org/10.1103/PhysRevD.82.054021}{\doi{10.1103/PhysRevD.82.054021}},
\href{http://www.arXiv.org/abs/1004.4624}{\texttt{arXiv:1004.4624}}.
%%CITATION = ARXIV:1004.4624;%%.

\bibitem{CLs1}
\hrefCMSnoop {}{A.~L. Read, ``Presentation of search results: the {$CL_s$}
  technique'',} \textit{ J. Phys. G} \textbf{ 28} (2002) 2693,
\href{http://dx.doi.org/10.1088/0954-3899/28/10/313}{\doi{10.1088/0954-3899/28/10/313}}.
%%CITATION = JPHGB,G28,2693;%%.

\bibitem{Junk:1999kv}
\hrefCMSnoop {}{T.~Junk, ``Confidence level computation for combining searches
  with small statistics'',} \textit{ Nucl. Instrum. Meth. A} \textbf{ 434}
  (1999) 435,
  \href{http://dx.doi.org/10.1016/S0168-9002(99)00498-2}{\doi{10.1016/S0168-9002(99)00498-2}},
\href{http://www.arXiv.org/abs/hep-ex/9902006}{\texttt{arXiv:hep-ex/9902006}}.
%%CITATION = HEP-EX/9902006;%%.

\bibitem{Eboli:2006wa}
\hrefCMSnoop {}{O.~J.~P. {\'{E}}boli, M.~C. Gonzalez-Garcia, and J.~K.
  Mizukoshi, ``$pp\rightarrow jj e^\pm \mu^\pm \nu\nu$ and
  $jje^\pm\mu^\mp\nu\nu$ at {\cal o}($\alpha^6_{\rm em}$) and {\cal
  o}($\alpha_{em}^4\, \alpha_s^2$) for the study of the quartic electroweak
  gauge boson vertex at {CERN LHC}'',} \textit{ Phys. Rev. D} \textbf{ 74}
  (2006) 073005,
  \href{http://dx.doi.org/10.1103/PhysRevD.74.073005}{\doi{10.1103/PhysRevD.74.073005}},
\href{http://www.arXiv.org/abs/hep-ph/0606118}{\texttt{arXiv:hep-ph/0606118}}.
%%CITATION = HEP-PH/0606118;%%.

\bibitem{1943Wald:wilks1938}
\hrefCMSnoop {}{S.~S. Wilks, ``The large-sample distribution of the likelihood
  ratio for testing composite hypotheses'',} \textit{ Ann. Math. Statist.}
  \textbf{ 9} (1938) 60,
  \href{http://dx.doi.org/10.1214/aoms/1177732360}{\doi{10.1214/aoms/1177732360}}.

\bibitem{Gounaris:1993fh}
\hrefCMSnoop {}{G.~J. Gounaris, J.~Layssac, and F.~M. Renard, ``Unitarity
  constraints for transverse gauge bosons at {LEP} and supercolliders'',}
  \textit{ Phys. Lett. B} \textbf{ 332} (1994) 146,
  \href{http://dx.doi.org/10.1016/0370-2693(94)90872-9}{\doi{10.1016/0370-2693(94)90872-9}},
\href{http://www.arXiv.org/abs/hep-ph/9311370}{\texttt{arXiv:hep-ph/9311370}}.
%%CITATION = HEP-PH/9311370;%%.

\bibitem{Baglio:2014uba}
J.~Baglio\hrefCMSnoop {}{ {et~al.}, ``{Release Note ---- VBFNLO 2.7.0}'',}
  (2014).
\href{http://www.arXiv.org/abs/1404.3940}{\texttt{arXiv:1404.3940}}.
%%CITATION = ARXIV:1404.3940;%%.

\end{thebibliography}\endgroup
\cleardoublepage \appendix\section{The CMS Collaboration \label{app:collab}}\begin{sloppypar}\hyphenpenalty=5000\widowpenalty=500\clubpenalty=5000\textbf{Yerevan Physics Institute,  Yerevan,  Armenia}\\*[0pt]
V.~Khachatryan, A.M.~Sirunyan, A.~Tumasyan
\vskip\cmsinstskip
\textbf{Institut f\"{u}r Hochenergiephysik,  Wien,  Austria}\\*[0pt]
W.~Adam, E.~Asilar, T.~Bergauer, J.~Brandstetter, E.~Brondolin, M.~Dragicevic, J.~Er\"{o}, M.~Flechl, M.~Friedl, R.~Fr\"{u}hwirth\cmsAuthorMark{1}, V.M.~Ghete, C.~Hartl, N.~H\"{o}rmann, J.~Hrubec, M.~Jeitler\cmsAuthorMark{1}, A.~K\"{o}nig, I.~Kr\"{a}tschmer, D.~Liko, T.~Matsushita, I.~Mikulec, D.~Rabady, N.~Rad, B.~Rahbaran, H.~Rohringer, J.~Schieck\cmsAuthorMark{1}, J.~Strauss, W.~Treberer-Treberspurg, W.~Waltenberger, C.-E.~Wulz\cmsAuthorMark{1}
% ...
\vskip\cmsinstskip
% ...
1:~~Also at Vienna University of Technology, Vienna, Austria\\
% ...

\end{sloppypar}
\end{document}